\documentstyle[aps,prb,eqsecnum,epsf]{revtex}
\begin{document}

\title{On the spin density wave transition in a two dimensional spin liquid}
\bigskip
\author{B. L. Altshuler}
\address{NEC Research Institute, 4 Independence Way, Princeton, NJ 08540 \\
and Physics Department, MIT, Cambridge, MA 02139}
\author{L. B. Ioffe}
\address{Physics Department, Rutgers University, Piscataway, NJ 08855 \\
and Landau Institute for Theoretical Physics, Moscow}
\author{A. I. Larkin}
\address{AT\&T Bell Laboratories, Murray Hill, NJ 07974 \\
and Landau Institute for Theoretical Physics, Moscow}
\author{A. J. Millis}
\address{AT\&T Bell Laboratories, Murray Hill, NJ 07974}

\maketitle

\begin{abstract}

Strongly correlated two dimensional electrons are believed to form a spin
liquid in some regimes of density and temperature.
As the density is varied, one expects a transition from this spin liquid state
to a spin density wave antiferromagnetic
state.
In this paper we show that it is self-consistent to assume that this transition
is second order and, on this assumption, determine
the critical behavior of the $2p_F$ susceptibility,  the NMR rates
$T_1$ and $T_2$ and the uniform susceptibility.
We compare our results to data on high $T_c$ materials.

\end{abstract}

\section{Introduction}

High temperature superconductors may be created adding carriers to  magnetic
insulators. At low dopings the compounds have long range magnetic order
at $T=0$; at high doping they do not.  Therefore, a
$T=0$ magnetic-non-magnetic transition must occur when the carrier density
exceeds a critical value.
The idea that some of the anomalous properties of high $T_c$ superconductors
are due to their proximity to this quantum phase transition has attracted
substantial recent interest.

The properties of the transition depend on the ordering wavevector and on the
nature of the disordered phase.
Several different possibilities have been studied in some detail including
antiferromagnet-singlet transitions in insulating magnets \cite{Sachdev}
and ferromagnetic and antiferromagnetic transitions in Fermi liquids
\cite{Hertz,Millis}.
Here we consider an important case which has not so far been discussed in the
literature, namely, that the disordered phase is a ``spin liquid'' and the
ordering occurs at the wavevector $|{\bf Q}| =2p_F$.

By ``spin liquid'' we mean a liquid of charge zero spin $1/2$ fermion
excitations coupled by a singular gauge field interaction; in the ground state
the fermions fill a large Fermi sea which occupies a substantial part of the
Brillouin zone \cite{spin-liquid}.
The spin liquid model has been argued to describe the normal phase of high
temperature superconductors \cite{Lee}.
By ``$2p_F$'' we mean a wavevector which connects two points on the Fermi
surface with parallel tangents (See Fig. 1) \cite{remark}.
For a circular Fermi surface any vector $\bf Q$ of magnitude $2p_F$ connects
two such points.
One important motivation for studying the $2p_F$ case is the high $T_c$
superconducting material $La_{2-x} Sr_x CuO_4$, in which strong magnetic
fluctuations have been observed \cite{Aeppli}; the fluctuations are peaked at
an $x$-dependent wavevector ${\bf Q}(x)$ which is claimed to be a ``$2p_F$''
wavevector of the Fermi surface calculated by standard band-structure
techniques for this material \cite{Littlewood}.

The $2p_F$ spin density wave transition in a spin liquid is different from the
$2p_F$ transition in a conventional Fermi liquid because the gauge field
interaction leads to divergences in the fermion response functions at
wavevectors $|{\bf Q}|$ near $2p_F$.
In this paper we calculate the exponents and the scaling functions
characterizing the $2p_F$ transition in the spin liquid.
Our starting point is a Hamiltonian, $H=H_{FL}+H_{gauge}$, where $H_{FL}$
describes fermions moving in a lattice and interacting with each other via a
short range four fermion interaction $W$:
\begin{equation}
H_{FL} = \sum_{p \alpha} \epsilon(p) c^\dagger_{p,\alpha} c_{p,\alpha} +
	W \sum_{p,p',q,\alpha,\beta} c^\dagger_{p,\alpha} c_{p+q,\alpha}
		c^\dagger_{p',\beta} c_{p'-q,\beta}
\label{H}
\end{equation}
and $H_{gauge}$ describes the gauge field and its coupling to the fermions;
it is discussed in detail in the literature \cite{spin-liquid,AIM} and below in
Section II.

\begin{figure}
\centerline{\epsfxsize=6cm \epsfbox{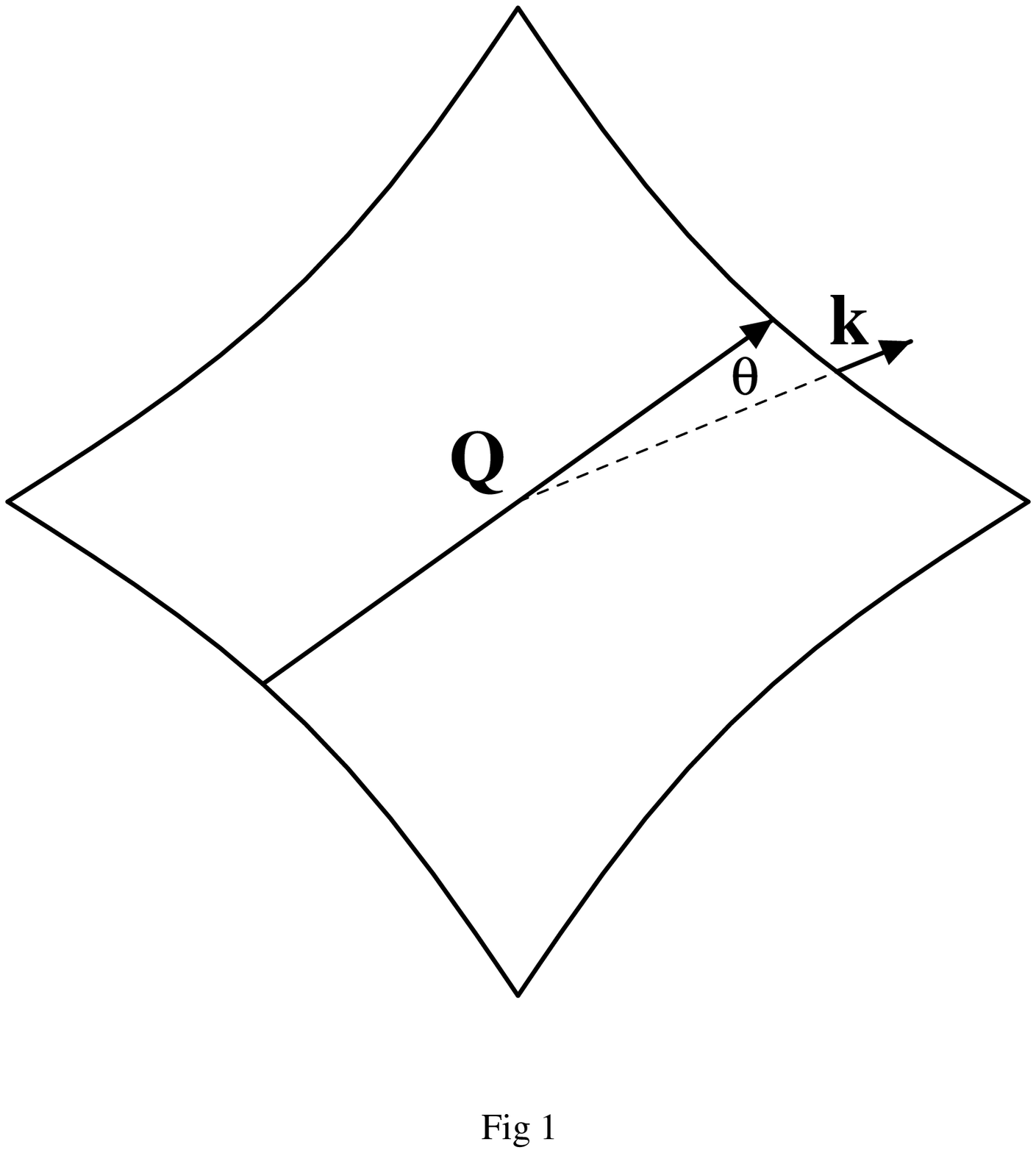}}
\caption{
Sketch of Fermi surface and important wavevectors.  The Fermi surface shown
here is similar to that claimed  to be appropriate to
$La_{1.8} Sr_{.14} Cu O_4$ .  The vector $\bf Q$ connects two points on the
Fermi surface.  It is assumed that the tangent to the Fermi surface at one end
of the vector $\bf Q$ is parallel to the tangent to the Fermi surface at the
other end.  We parametrize the vector $\bf k$ by the angle $\theta$ shown on
the figure and the magnitude $k_{\parallel}$ shown on the figure as the solid
line part of $\bf k$.
}
\end{figure}

It has recently been shown that two possibilities arise, depending on the
strength of the fermion-gauge-field interaction \cite{AIM}.
If this interaction is weak, a critical value, $W_c$, of the short-range
four-fermion interaction separates a disordered phase with  non-divergent spin
fluctuations at $W <W_c$ from an ordered phase at $W > W_c$.
In this paper we evaluate the exponents characterizing the transition at
$W = W_c$ in the weak gauge-field coupling case and show how
they depend on the strength of the fermion-gauge-field
interaction.
If, however, the fermion gauge-field interaction is sufficiently strong, then
the $2p_F$ spin susceptibility diverges as $T \rightarrow 0$ for arbitrary
$W<W_c$, although there is no long range order at $T=0$ \cite{AIM}.
We have not studied the transition at $W = W_c$ in the strong gauge-field
interaction case, but we give the exponents characterizing the $W <W_c$ phase
and discuss the physical consequences.

We assume that transition at $W=W_c$ leads to a spin density wave state with
long range order with wave vector $\bf Q$ and that this transition is of a
second order.
If $|{\bf Q}| \neq 2p_F$, the gauge field does not modify the fermion
susceptibility \cite{AIM}, so we expect previously developed theory
\cite{Hertz,Millis} to apply.
We now outline our approach to the $|{\bf Q}| = 2p_F$ case.
Because we expect the physics in this region to be determined by the exchange
of spin density fluctuations we use a Hubbard-Stratonovich transformation to
recast Eq. (\ref{H}) as a theory of fermions coupled to spin fluctuations $\bf
S_{q}$:
\begin{eqnarray}
H' &=& \sum_{p \alpha} \epsilon(p) c^\dagger_{p,\alpha} c_{p,\alpha} +
	g \sum_{p,k,\alpha,\beta} c^\dagger_{p,\alpha}
	\vec{\sigma}_{\alpha \beta} c_{p+k,\beta} \vec{S_{-k}}
\label{H'} \\
&& +	H_{gauge} + \sum_{k}  {\bf S}_k^2
\nonumber
\end{eqnarray}
Here $g$ is a fermion-spin fluctuation coupling constant derived from $W$; in
weak coupling $g^2=W$.
One may study $H'$ as it stands or one may integrate out the fermions
completely, obtaining a theory of interacting spin fluctuations which is
described by the action
\begin{eqnarray}
{\cal A}[S] &=& \sum_{k, \omega} \vec{S}_{\omega , k} \chi_0^{-1}(k, \omega )
	\vec{S}_{- \omega , k}
\label{A[S]} \\
&+& \frac{1}{4}
	\sum_{\omega_i , k_i} U_{\omega_1 , ... \omega_4}^{k_1,...k_4}
	( \vec{S}_{\omega_1, k_1} \cdot \vec{S}_{\omega_2, k_2})
	( \vec{S}_{\omega_3,k_3} \cdot \vec{S}_{\omega_4,k_4})
	\delta ( \sum_i  \omega_i) \delta ( \Sigma_i k_i) + \ldots
\nonumber
\end{eqnarray}
Here $\vec{S}_{\omega , k}$ represents a spin fluctuation of Matsubara
frequency $\omega$ and wavevector $\bf k$.
$\chi_0$ is the susceptibility and $U$ is a four-spin-fluctuation interaction
proportional to $g^4$.
We shall derive and interpret this action in more detail below.
This action is difficult to treat even for a Fermi liquid without the gauge
field interaction because $\chi_0$ and $U$ diverge as $T,\omega \rightarrow 0$
and ${\bf
k} \rightarrow {\bf Q}$.
The gauge field causes additional singular renormalizations of
$\chi_0$, $U$, $g$ and the fermion propagator \cite{AIM,IoffeKivelson}.

In this paper we present and justify a a self-consistent one-loop
approximation  method for extracting physical results from the formally
divergent theory.
We supplement this treatment by an analysis based on Eq. (\ref{H'}) of the
effect of the spin fluctuations on the fermions.
We find that the susceptibility is less singular than the
susceptibility obtained for transitions with $Q \neq 2p_F$.
This is because the $2p_F$ singularity of the fermions leads to a long ranged
(RKKY) interactions;
this weakens the singularities associated with the transition for the same
reason that long ranged dipolar interactions lower the critical dimension
of classical models of ferroelectric transition \cite{larkinkhmelnitskii}.

The self-consistent one loop approximation succeeds in the spin liquid case
because the fermion-gauge field interaction makes the singular part of the
fermion response symmetric in the variable $|{\bf k}| - 2 p_F$.
This approach however fails for the conventional Fermi liquid  because
in this case the
singularities in the fermion polarizability are not symmetric in the variable
$|{\bf k}| - 2p_F$ as discussed below.
We have found that this asymmetry implies that fluctuation effects do
not permit a second order transition
at $|{\bf Q}|=2p_F$ in a Fermi liquid.
We will present a detailed treatment elsewhere \cite{2pf}.

The outline of this paper is as follows.
Section II reviews the relevant theory of the spin liquid.
Section III is devoted to the transition in the spin liquid for weak fermion
gauge-field interaction.
Section IV discusses the properties of the small-$W$ critical phase occurring
when the gauge interaction is strong enough that the $2p_F$ susceptibility
diverges as $T \rightarrow 0$.
Readers uninterested in the details of the derivations may proceed directly to
Section V which contains a summary of the results and a conclusion.

\section{Gauge theory of spin liquid}

In this section we review the properties of the ``spin liquid'' regime of the
$t$-$J$ model, which has been argued to describe the normal state of the
high-$T_c$  superconductors \cite{Lee}.
In the spin liquid regime the elementary excitations are charge $e$, spin $0$
``holons'', charge $0$, spin $1/2$ fermionic ``spinons'', and a gauge field
which mediates a long ranged, singular interaction.
For temperatures above the Bose condensation temperature of the holons, the
holons have a  negligible effect on the magnetic properties, which are
determined by the properties of the degenerate Fermi gas of spinons (which we
call a spin-liquid),  coupled by the gauge interaction and
by a short-range four-fermion interaction $W$.

The properties of the spin liquid have  been studied by many authors
\cite{spin-liquid,Lee,AIM}.
The fermion propagator has been found to have the form
\begin{equation}
G = \frac{1}
	{i \omega_0^{1/3} \omega^{2/3} - v (p_{\parallel} +p_{\perp}^2/p_0)}
\label{G}
\end{equation}
instead of the Fermi liquid form
$G= [i\omega - v (p_{\parallel} + p_{\perp}^2 /p_0) ]^{-1}$.
Here $v_F$ is the Fermi velocity, $p_0$ is the radius of curvature of the
Fermi line and $p_{||} (p_{\perp} )$ are momentum components normal
(tangential) to the Fermi line as measured from the points $\pm {\bf Q}/2$ and
$\omega_0$ is an energy scale of order $p_F v_F$ which is
defined more precisely in Ref. \cite{AIM}.

\begin{figure}
\centerline{\epsfxsize=6cm \epsfbox{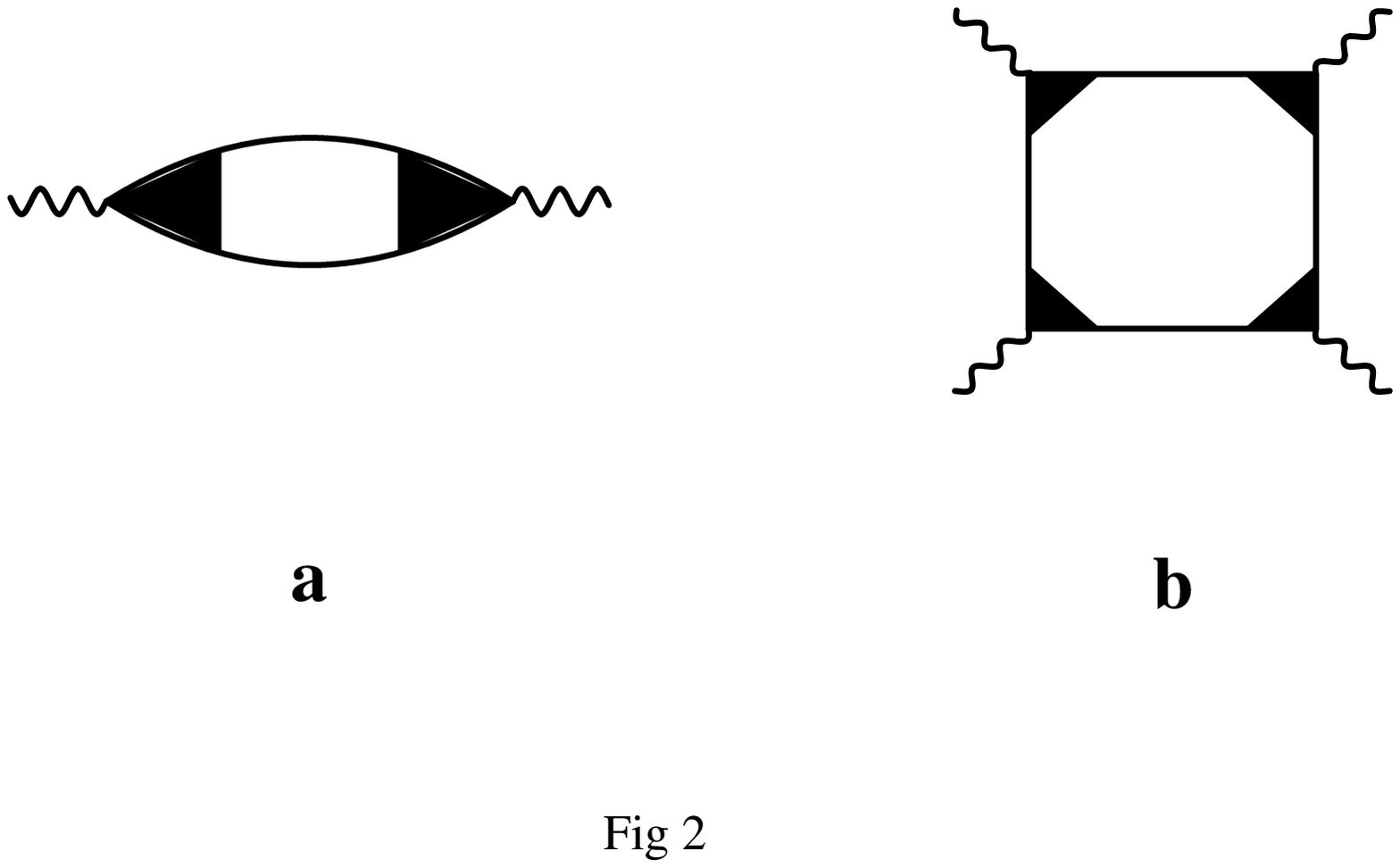}}
\caption{
(a)  Diagram yielding the nonanalytic momentum and frequency dependence
of the susceptibility $\chi_0$.  The solid lines are the fermion propagators.
The shaded triangles are vertices which have power law divergences.
(b)  Diagram yielding the nonanalytic momentum and frequency dependence
of the four spin fluctuation interaction $U$.
}
\end{figure}

Recently, we have shown \cite{AIM} that the fermion-gauge-field interaction
causes a
power law divergence in the vertex $\Gamma_{2p_F}$ which is shown as a black
triangle in Fig. 2a.
Specifically we found,
\begin{equation}
\Gamma_{2p_F} \simeq \left[ \frac{1}{
	\max\left[ \left(\frac{\omega}{\omega_0}\right),
		\left(\frac{v_F k_{\parallel}}{\omega_0}\right)^{3/2} \right]}
	\right]^{\sigma}
\label{Gamma_2pF}
\end{equation}
We were able to calculate $\sigma$ in the limits $N \rightarrow
\infty$ and $N \rightarrow 0$, where $N$ is the spin degeneracy of the
fermions.
We found
\begin{eqnarray*}
\sigma &\sim 1/2N \hspace{1in} & N \rightarrow \infty \\
\sigma &\sim 16 \sqrt{2} /9 \pi \sqrt{N} \hspace{0.2in} & N \rightarrow 0.
\end{eqnarray*}
Extrapolation of these results to the physical case $N=2$ gives the estimates
$1/4 \lesssim \sigma \lesssim 3/4$.
The combination of the change in form of the Green function and the divergence
of  $\Gamma_{2p_F}$ has profound effects on the polarization operator
$\Pi(\omega,k)$  (which is shown in Fig. 2a) and in the four spin fluctuation
interaction $U$  (which is shown in Fig. 2b, cf. Eq. (\ref{U})).

If $1/6 < \sigma < 1/3$, we found \cite{AIM}
\begin{equation}
\Pi ( \omega , k_{\parallel} ) = \Pi_0 - \sqrt{\frac{\omega_0 p_0}{v_F^3}}
	\left[ c_{\omega} \left (\frac{|\omega|}{\omega_0}\right)^{\frac{2}{3}
		-2\sigma}
	+ c_k Re \left(\frac{|k_{\parallel}|v_F}{\omega_0} \right)^{1-3\sigma}
	\right]
\label{Pi1}
\end{equation}
Here $c_{\omega}$ and $c_k$ are constants of order unity.
For $\sigma < 1/6$, the term involving $k_{\parallel}$ has the same sign as
$k_{\parallel}$.
Thus the maximum of $\Pi$ and therefore the ordering wavevector must
occur at a $|{\bf Q}|$ different from $2p_F$ and we expect previous treatments
\cite{Hertz,Millis} of the critical behavior to apply.
For $\sigma > 1/6$ the $k_{\parallel}$ term is always positive and a second
order transition at $|{\bf Q}|=2p_F$ is possible.
We restrict our attention to $\sigma > 1/6$ in the rest of the paper.
In this case the nonanalyticity of  $Re k_{\parallel}^{1-3\sigma}$ affects
only the magnitude of $c_k$, that is not important for the scaling arguments
which we will present.
We henceforth write the   $k_{\parallel}$ term as
\[
c_k \left(\frac{k_{\parallel} v_F}{\omega_0} \right)^{1-3\sigma}.
\]

On the other hand, if $\sigma > 1/3$, $\Pi$ diverges as
$\omega , k_{\parallel} \rightarrow 0$.
We found \cite{AIM}
\begin{equation}
\Pi ( \omega , k_{\parallel} ) = \sqrt{\frac{\omega_0 p_0}{v_F^3}}
	\frac{1}{
	\left [ c_{\omega}
	\left ( \frac{|\omega|}{\omega_0} \right )^{2\sigma - 2/3}
	+  c_k \left(\frac{|k_{\parallel}| v_F}{\omega_0} \right)^{3\sigma-1}
	\right ] }
\label{Pi2}
\end{equation}

The full susceptibility $\chi$ is obtained by combining the irreducible bubble
$\Pi$ with the short-range four fermion vertex $W$.
We have shown \cite{AIM} that the gauge-field interaction renormalizes a
sufficiently weak initial $W_i$ to zero.
Therefore we expect a $T=0$ transition as $W$ is varied through a critical
value $W_c$.
In the weak coupling case $\sigma < 1/3$ the fact that $\Pi(\omega =0,
k_{\parallel} = 0)$ is non-divergent implies that the usual RPA formula
\begin{equation}
\chi(\omega,k) = \frac{\Pi(\omega,k)}{1- W \Pi(\omega,k) }
\label{chi_RPA}
\end{equation}
is the correct starting point of a theory of the transition.
We discuss the transition occurring when $W = W_c = \Pi (0,0)^{-1}$ in more
detail below in the next section.

In the strong coupling, $\sigma > 1/3$, case we still expect a transition when
$W$ exceeds a critical value.
However the divergence of $\Pi (0,0)$ implies that the RPA formula
(\ref{chi_RPA}) is not correct.
For $W < W_c$, $\chi(\omega,k) = \Pi(\omega,k)$ with corrections of order
the product of $W$ and a positive power of frequency or $k_{\parallel}$.
The $T=0$ critical point separates a $W<W_c$ phase which has
power-law spin correlations from a $W>W_c$ phase which has long-range order.

\section{Weak fermion-gauge-field coupling}

There are two possible approaches to the spin density wave transition at
$W=W_c$.
One is to integrate out the fermions, obtaining an effective action,
${\cal A}[S]$,  describing the spin fluctuations, $S_{\omega ,k}$, and then
to analyze this action.
If $Q \neq 2p_F$ all terms in the effective action are finite in both the
Fermi liquid and in the spin liquid, and the action may be treated by standard
renormalization group  techniques \cite{Millis}.
As $Q \rightarrow 2p_F$, however, the coefficients of the quartic and higher
order terms diverge, implying that standard RG techniques cannot be used and
that the effect of higher order terms must be investigated.
Another approach is to investigate the effect of the spin fluctuations on the
fermion propagator.
Here we consider both approaches.

Taking the first approach we proceed in four steps.
First, we integrate out the fermions obtaining the action ${\cal A}[S]$ given
in Eq. (\ref{A[S]}).
Second, we truncate the action, retaining only the quadratic and quartic terms.
Third, we solve the truncated action in the self-consistent one loop
approximation.
This approximation has been extensively used to study three dimensional
magnets \cite{Moriya} and gives the same results as the renormalization
group above the upper critical dimension \cite{Millis}.
Fourth, we show using the formalism of the effective action ${\cal A}[S]$
that corrections to the self-consistent one loop approximation lead at most
to logarithmic corrections to the self consistent one loop results.
Fifth, we confirm this result using the formalism of the spin fluctuations
interacting
with fermions. Finally, we discuss the physical consequences of the theory.

\subsection{Derivation of truncated action}

We begin by evaluating the coefficients $\chi_0$ and $U$ in action
(\ref{A[S]}).
We shall be interested in momenta close to the momentum $\bf Q$ at which
$\chi_0$ is maximal.
For wavevectors near $\bf Q$ the momentum and frequency dependence of $\chi$
and $U$ are non-analytic and controlled by Fermi surface singularities.
It will be convenient to parametrize the momentum $\bf k$ in terms of a
magnitude $k_{\parallel}$ and an angle $\theta$ as shown in Fig. 1.
For each angle $\theta$  there is a momentum ${\bf p}_F ( \theta )$ such that
$2 {\bf p}_F(\theta )$ spans the Fermi surface.
Note $2{\bf p}_F(\theta = 0) = {\bf Q}$.
We define $k_{\parallel}$ to be the difference $|{\bf k} |-2 p_F(\theta )$.
These definitions are generalizations for the noncircular Fermi surface of
coordinates which are the convenient choice for a circular Fermi surface.

The fermion contribution $\Pi(\omega,k_\parallel)$ to the inverse
susceptibility $\chi_0^{-1}$  can be calculated by  summing all diagrams which
are irreducible with respect to the fermion-fermion interaction and have two
external $S_{\omega, k}$ legs.
This sum has contributions from short length scale processes which give
$\Pi(\omega,k_\parallel)$ an analytic dependence on $k_{||}$ and $\theta$ and
also contributions from Fermi surface singularities, which lead to a
nonanalytic dependence of $\Pi(\omega,k_\parallel)$ on $k_{||}$ and $\omega$.
The Fermi surface singularities come from the diagram shown in Fig. 2a
which leads to the expression (\ref{Pi1}).
Thus we get the susceptibility

\begin{equation}
\chi_0(\omega,k) \simeq \frac{\sqrt{\frac{v_F^3}{\omega_0 p_0}}}
	{g^4  \left[c_{\omega} \left(
		\frac{|\omega|}{\omega_0}\right)^{2/3-2\sigma}
	+ c_k \left(\frac{|k_{\parallel}| v_F}{\omega_0} \right)^{1-3\sigma}
	+ \theta^2
	+ \left( \frac{\kappa_0}{p_F} \right)^2 \right] }
\label{chi_0}
\end{equation}
where $g$ is the fermion-spin fluctuation coupling.
The coefficients $c_\omega$ and $c_k$ are of the order of unity, they are
sensitive to the details of the band structure and the momentum dependence of
the interaction.  The coefficient $\kappa_0^2$ is
determined by difference of the interaction from its critical value.
Note that we are using Matsubara frequencies so that $\chi$ is purely real.

The most singular contribution, $U_{sing}$, to $U$ is given by the diagram
shown in Fig. 2b.
It diverges if the reduced momenta $( k_{||} , \theta )$ and frequencies of
all four legs are zero.
For our subsequent calculations we shall need to estimate the asymptotic
behavior when the momenta and frequencies on the external legs satisfy
\begin{eqnarray*}
k_{\parallel}  \approx k_{\parallel a} \approx k_{\parallel b} &\gg&
	k_{\parallel c}  \approx k_{\parallel d}
\\
\omega \approx \omega_a \approx \omega_b &\gg& \omega_c \approx \omega_d
\end{eqnarray*}
and $\theta_a \approx \theta_b$; $\theta_c \approx \theta_d$ but with
$\theta_a - \theta_c$ arbitrary.
Here a,b,c,d are any permutation of the legs shown in Fig 2b.
By evaluating the diagram in Fig. 2b we find that in this limit the vertex is

\begin{equation}
U(k_{\parallel} , \omega ; \theta_1,\theta_2) \simeq
	\frac{g^4}{\omega_0^2} \sqrt{\frac{\omega_0 p_0}{v_F^3}}
	\frac{1}
	{ \left ( \frac{|\omega|}{\omega_0} \right )^{4\sigma+2/3} +
	\left(\frac{v_F |k_{\parallel}|}{\omega_0} \right)^{6\sigma+1} +
	\left(\frac{v_Fp_F^2(\theta_1 -\theta_2)^2}{p_0 \omega_0}
		\right)^{6 \sigma+1} }
\label{U}
\end{equation}
The equation (\ref{U}) is not strictly correct at arbitrary momentum and
frequency but gives the correct asymptotics of the vertex, which are all we
need.
The result may be motivated by the following arguments:
the scaling $\omega_0^{1/3}\omega^{2/3} \sim v_F k_{\parallel}
\sim v_F (p_F \theta )^2 /p_0$
comes from the structure of the expanded fermion Green function (\ref{G});
we believe that this is the correct generic scaling in this problem.
In the case of the circular Fermi surface the dependence
on the large angular separation
($\theta_a - \theta_c $) follows from rotational invariance; a noncircular
Fermi surface can still be mapped locally to a circle, proving
this result in the general case.
In writing these equations we have included the renormalization of
$\Gamma_{2p_F}$ due to the fermion-gauge-field interaction.
The renormalization of the fermion vertex $g^2$ due to the gauge field
does not enter our discussion of the critical phenomena.
To see this, note that the renormalization is due to the logarithmic divergence
of the diagram shown in Fig. 3.
In the present case the interaction is sharply peaked, so the logarithm is
confined to momenta and frequences less than the  scale set by $\kappa$ (see
Eq. (\ref{chi_0})), and therefore does not affect our treatment of the
critical point.
The problem is thus to investigate the self-consistency of the
theory defined by Eqs. (\ref{A[S]}), (\ref{chi_0}) and (\ref{U}).

\begin{figure}
\centerline{\epsfxsize=6cm \epsfbox{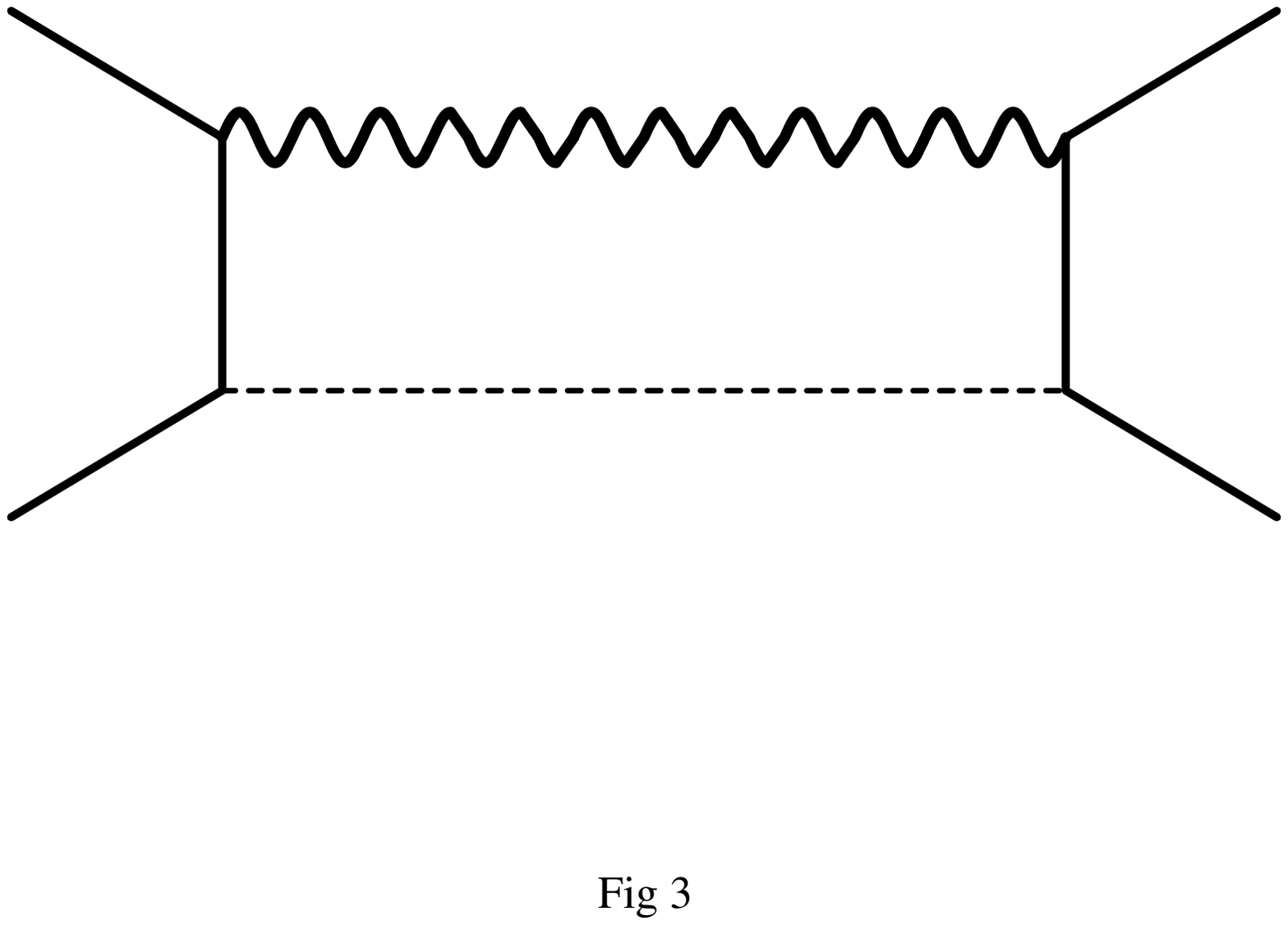}}
\caption{
Diagram yielding renormalization of the spin-fluctuation contriubtion to the
four fermion vertex $W$ by gauge field.  Here the solid lines are fermions, the
dashed line represents the gauge field, and the wavy line represents the spin.
}
\end{figure}

\subsection{Self-consistent one-loop approximation}

We treat the theory defined by Eqs. (\ref{A[S]},\ref{chi_0},\ref{U}) in
the self-consistent one loop approximation shown diagrammatically in Fig. 4,
i.e. we require that  the full susceptibility obeys the equation
\begin{equation}
\chi(k, \omega )^{-1} = \chi_0^{-1} ( k, \omega ) +
g^4 \sum_{\eta ,q} \chi (\eta ,q) U_{\eta,\omega}^{k,q}
\label{chi_eq}
\end{equation}
where $\chi_0$ is given by (\ref{chi_0}).
Using (\ref{chi_0}) and (\ref{U}) we see that the frequency or momentum
derivative of the integral (\ref{chi_eq}) is infra-red divergent.
Estimating this integral we get:
\begin{equation}
\chi^{-1} (k, \omega) - \chi_0^{-1} (k, \omega) = g^4
	\sqrt{\frac{\omega_0 p_0}{v_F^3}}
	\left [ \left ( \frac{| \omega |}
	{\omega_0} \right )^{2/3 - 2 \sigma}
 	+ \left (\frac{v_F|k_\parallel|}{\omega_0} \right )^{1-3\sigma}\right ]
\end{equation}
which has the same order of magnitude and the same frequency
and momentum dependence as the bare $\chi_0^{-1}$.
Here as in Eq. (\ref{U}) the formula gives only the correct asymptotics when
one of the variables ($\omega_0^{1/3}\omega^{2/3} , v_F k_{||}$,
$v_F(p_F \theta )^2 / p_0$,
$v_F \kappa^2 / p_0 )$ is much larger than the others.

\begin{figure}
\centerline{\epsfxsize=6cm \epsfbox{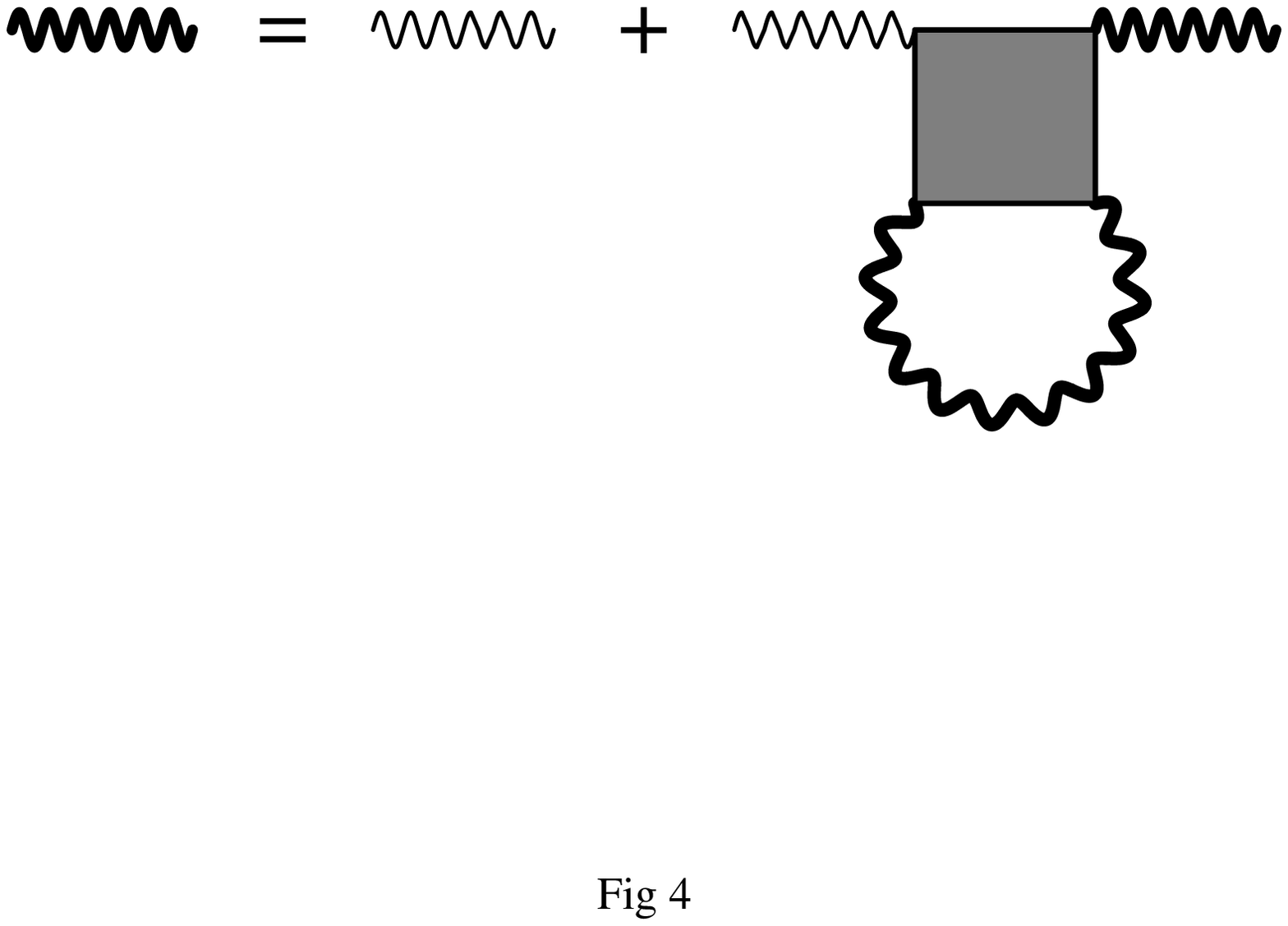}}
\caption{
Diagrams defining the self-consistent one-loop approximation for the
susceptibility $\chi$ (thick wavy line) in terms of the interaction $U$ (shaded
box) and the bare susceptibility $\chi_0$ (thin wavy line).
}
\end{figure}

Due to the four spin interaction $U$ the cutoff, $\kappa_R^2 (T)$, is
renormalized as compared to $\kappa_0^2$.
At a finite temperature the sum over $\eta$ in (\ref{chi_eq}) is taken
over Matsubara frequencies $\eta_n =2 \pi n T$.
The temperature dependent part of the correction is dominated by the first
term in this sum.
Estimating it at zero external momenta and frequency we get the leading
contribution to
\begin{equation}
\frac{\kappa^2(T)}{p_F^2} =
	\left ( \frac{T}{\omega_0} \right )^{\frac{2}{3} - 2 \sigma}
	+ \frac{\kappa^2(0)}{p_F^2}
\label{kappa(T)}
\end{equation}

Note that $\chi^{-1}( \theta ) \sim \theta^2$; the singular interaction does
not change the power law in the angle dependence.
This may be most easily seen via a {\it reductio ad absurdum.}
Suppose the four-spin-wave interaction had led to an exponent less than two
for $\theta$.
Then the angular dependence of $\chi_0$ in Eq. (\ref{chi_eq}) could have been
neglected.
However for a circular Fermi surface Eqs. (\ref{U},\ref{chi_eq}) are
rotationally invariant (apart from terms due to $\chi_0$),
and can therefore lead to no angular dependence at all, in contradiction to
the hypothesis  of angular dependence with an anomalous exponent.
The same argument applies to a nonspherical Fermi surface,
because it may be mapped into a spherical one, with errors of order $\theta^2$.
We conclude that in general the self-consistent one-loop
equations cannot produce a $\theta$-dependence
different from $\theta^2$.

\subsection{Marginality of Higher Order Corrections}

We now argue that corrections to the self-consistent solution (\ref{chi_0})
do not change the momentum, frequency and temperature dependences of
Eqs. (\ref{chi_0},\ref{kappa(T)}).
There are two kinds of corrections: those arising from higher orders of
perturbation theory in the $S^4$ coupling using the truncated
action, Eq. (2.1), and those arising from higher order $S^6 , S^8 ...$
nonlinearities omitted from Eq. (\ref{A[S]}).
We first consider the renormalization of the vertex $U$
at the second order in $U$.
The corresponding diagram is shown in Fig. 5.
This diagram is infrared divergent and may therefore by evaluated by writing
the integrals with zero external momenta and frequencies and then cutting off
the resulting divergence by the largest of the external momenta or frequency.
This gives
\begin{eqnarray}
& & \delta U ( \Omega , q, \theta_1, \theta_2 ) \sim
	\frac{g^4 p_F}{\omega_0^4}
	\int_{\omega > \Omega} \int_{k_\parallel > q}
	\frac{d\omega  d \theta dk_{\parallel}}
	{\left[ \left(\frac{\omega}{\omega_0}\right)^{2/3-2\sigma} +
		\left(\frac{v_F k_{\parallel}}{\omega_0}\right)^{1-3\sigma}
		\right]^2 }
\nonumber \\
	&\times&
	\frac{1} { \left[
	\left ( \frac{|\omega|}{\omega_0} \right )^{4\sigma+2/3} \!\!+\!
	\left(\frac{v_F k_{\parallel}}{\omega_0} \right)^{6\sigma+1} \!\!+\!
	\left(\frac{v_Fp_F^2(\theta -\theta_2)^2}{p_0 \omega_0}
		\right)^{6 \sigma+1} \right]
	\left[
	\left ( \frac{|\omega|}{\omega_0} \right )^{4\sigma+2/3} \!\!+\!
	\left(\frac{v_F k_{\parallel}}{\omega_0} \right)^{6\sigma+1} \!\!+\!
	\left(\frac{v_Fp_F^2(\theta_1 -\theta)^2}{p_0 \omega_0}
		\right)^{6 \sigma+1} \right] }
\nonumber \\
&& \hspace{1in} \sim \frac{g^4}{\omega_0^2} \sqrt{\frac{\omega_0 p_0}{v_F}}
	\frac{1}{  \left( \frac{|\omega|}{\omega_0} \right )^{4\sigma+2/3} +
	\left(\frac{v_F k_{\parallel}}{\omega_0} \right)^{6\sigma+1} +
	\left(\frac{v_Fp_F^2(\theta_1 -\theta_2)^2}{p_0 \omega_0}
		\right)^{6 \sigma+1} }
\label{delta_U}
\end{eqnarray}
This estimate shows that higher order corrections in the $S_q^4$ interaction
do not change the power law dependence of the mode coupling interaction, and
therefore cannot change the powers coming from the solution
of the self-consistent equation.
This argument of course does not rule out logarithmic corrections to
$\chi^{-1}$.

\begin{figure}
\centerline{\epsfxsize=6cm \epsfbox{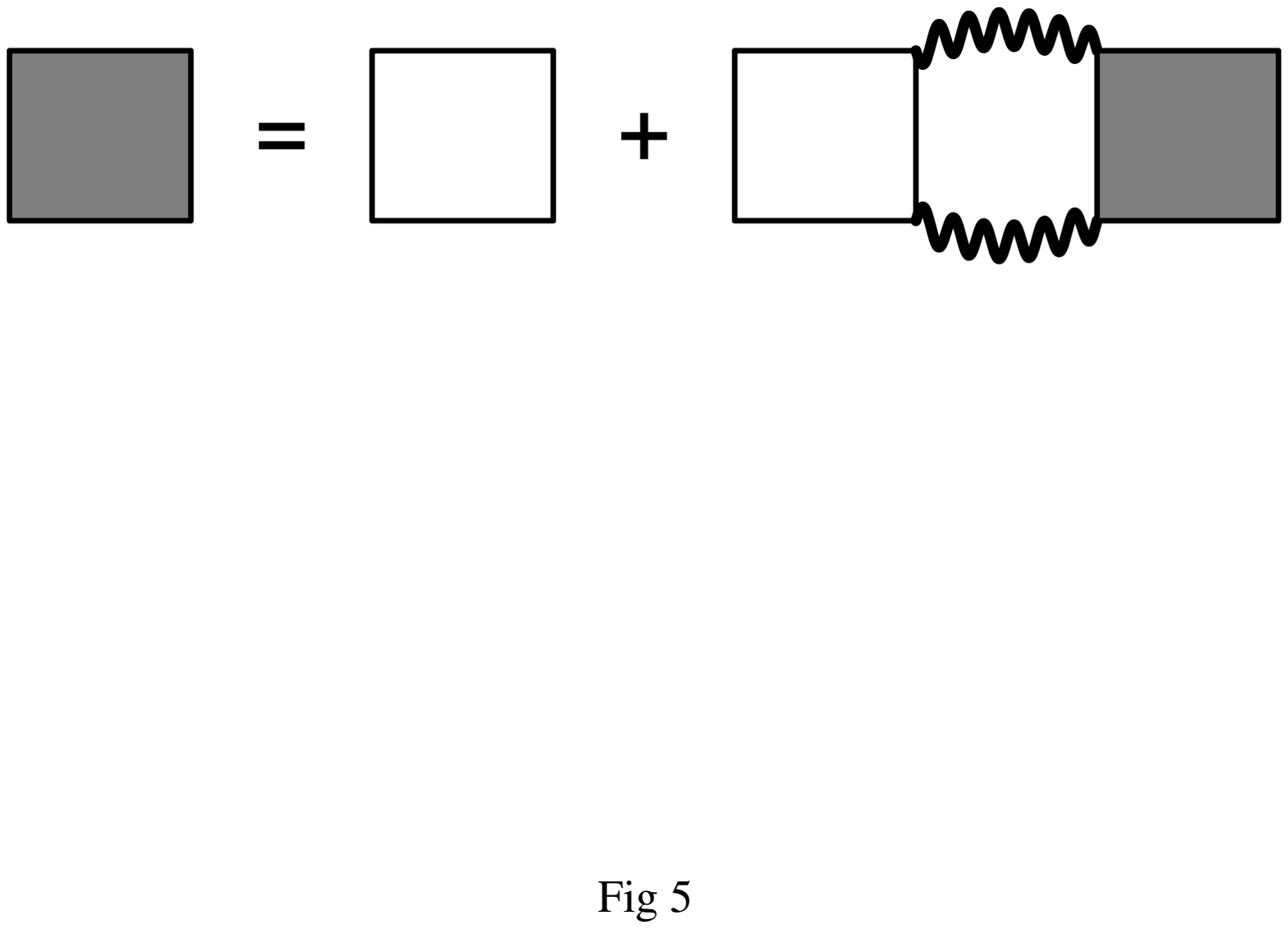}}
\caption{
Diagram giving leading renormalization of full four spin interaction $U$
(shaded box) in terms of bare $U$ (open box) and spin fluctuations (wavy
lines).
}
\end{figure}

The effect of the higher order nonlinearities
$(S_q^6, \; S_q^8$, etc) may be analyzed similarly.
Let us compare the contribution to the susceptibility from one bare
$S_q^{2n+2}$ vertex to the contribution from one bare $S_q^{2n}$ vertex.
The former contains one more integral over ($\omega , k_{||} , \theta$) and one
more factor of the susceptibility $\chi$.
The diagram for the bare vertex $S_q^{2n+2}$ contains also two more fermion
Green functions.
Performing the extra integration we see that the additional singularity coming
from the extra fermion Green functions in the bare vertex and extra factor of
susceptibility is precisely cancelled by the phase volume.

These results imply that the theory is marginal, in the sense that higher
order interactions give the same infrared behaviour as lower order
interactions.
The marginality may also be demonstrated by a scaling argument.
If we rescale momenta and frequencies via
$\omega^{\prime} = b \omega$,
$k_{\parallel}^{\prime} = b^{2/3}k_{\parallel}$,
$\theta^{\prime} = b^{1/3} \theta$, then we must scale the field $S_q^{\prime}
= b^{-4/3+\sigma} S_q$ to keep the quadratic term in the action invariant.
The scaling dimension of the $S_q^{2n}$ vertex is then
$b^{(-8/3+2\sigma)n}$ (from the fields) times
$b^{-4n/3+2-2n\sigma}$ (from the vertex) times
$b^{2(2n-1)}$ (from the integrals).
Adding the powers we see that the total scaling dimension of the vertex is
zero, so all interactions are marginal in the renormalization group sense.

Thus, we conclude that higher order effects of the spin wave interaction do
not change  the exponents characterizing the divergence of the susceptibility
(\ref{chi_0}) or the interaction between spin fluctuations (\ref{U}).

\subsection{Effect of spin fluctuations on fermions}

We now consider the  alternative approach of estimating the feedback of
the interaction mediated by the spin fluctuations (\ref{chi_0}) on the
fermions.
The lowest order contribution to the fermion self-energy $\Sigma$ is
\begin{equation}
\Sigma ( \epsilon , p) = g^2 \int G( \epsilon',p^{\prime}  )
\chi (\epsilon - \epsilon^{\prime}, p-p^{\prime} )
[ \Gamma_{(p+p^{\prime}) / 2 , ( \epsilon + \epsilon^{\prime} ) / 2}
( \epsilon - \epsilon^{\prime} ,p-p^{\prime})]^2
(dp^{\prime} d \epsilon^{\prime} )
\end{equation}

We are interested in the leading frequency dependence of $\Sigma$; the main
contribution to this comes when $p^\prime$ is on the Fermi line.
We then find the angle at which $\chi$ is peaked; this turns out to be the
point which is located symmetrically opposite to $p$:
$| \theta^{\prime} - \theta + \pi |
	\sim ( \kappa^2+4 \theta^2)^{\frac{1}{2(1-3 \sigma )}}$
where $\theta^{\prime}$ and $\theta$ are the polar
coordinates of the points $p^{\prime}$ and $p$.
Estimating the contribution of this region we get:
\begin{equation}
\Sigma ( \epsilon , \theta ) =
\frac{-ic_{\Sigma} \epsilon}{\left[ \left(\frac{\kappa}{p_F}\right)^2 +
	4\theta^2 \right]^{\frac{1}{2(1-3\sigma)}}}
\label{Sigma}
\end{equation}
where $c_{\Sigma} \sim 1$.
Over most of the Fermi line, this self-energy is small compared
to the fermion self-energy $\omega_0^{1/3} \epsilon^{2/3}$ due to the
gauge field interaction.
The spin fluctuation contribution becomes important only in a small region $|
\theta | \leq \theta^*$ near the points connected by the wave vector $Q$,
with $\theta^*$ given by
\begin{equation}
\theta^* = ( \epsilon / \omega_0 )^{\frac{(1-3 \sigma )}{3}}
\label{theta^*}
\end{equation}
At a finite temperature $T$ the typical energy of the fermion is
$T$, and it should replace $\epsilon$ in (\ref{theta^*}):
\begin{equation}
\theta^* (T) =
(T/ \omega_0 )^{\frac{(1-3 \sigma )}{3}}
\label{theta^*(T)}
\end{equation}
Since $\theta^*(T) \gtrsim \kappa/p_F$ (cf. Eq. (\ref{kappa(T)})),
in the region $\theta \lesssim \theta^*(T)$ the self-energy (\ref{Sigma})
remains of the order of $\omega_0^{1/3} \epsilon^{2/3}$.

We do not have a reliable expression for the self-energy $\Sigma ( \epsilon,
\theta )$ in the region $| \theta | < | \theta^* |$ at $\epsilon \gg T$.
We can, however, argue that fermions in this region
make only a marginal contribution to physical quantities such
as polarization operator $\Pi (\omega, k)$.
The analytic expression corresponding to Fig. 2a for $\Pi$ is
\begin{equation}
\Pi ( \omega, k_\parallel, \theta) = \int dp d\epsilon \Gamma_{2p_F}^2
G({\bf p+k}, \epsilon + \omega ) G({\bf p}, \epsilon)
\end{equation}
Here $\bf k$ is a vector parametrized as shown in Fig. 1 by variables
$k_\parallel, \theta$.
As a function of angular variable $\theta$, $\Pi(\omega, \theta)$ has a smooth
maximum of the width
$\theta'(\omega) \sim (\omega/\omega_0)^{1/3-\sigma}
\sim \theta^*(\omega)$.
Thus all physical quantities are controlled by spin fluctuations with
$\theta \gtrsim \theta'(\omega)$.
Using Eq. (\ref{G}) for $G$ and (\ref{Gamma_2pF}) for $\Gamma_{2p_F}$
we see that the integral over momenta is dominated by fermions with momenta
near the Fermi surface which have angles very close to $\theta$ (the dominant
contribution comes from the range of fermion angles $\theta'$:
$|\theta'-\theta| \sim (\omega/\omega_0)^{1/3} \ll \theta$).
Thus for $k_\parallel$ and $\theta$ which are important for the
susceptibility, for the self-energy $\Sigma(\epsilon)$ or for the
$T$-dependence of $\kappa$, the polarization bubble is controlled by the
fermions at $\theta \gtrsim \theta^*$.
Thus, we believe that the fermion-spin-wave interaction leads only to
corrections of the order of unity to physical quantities.
Certainly, these arguments are based on power counting and can easily miss
logarithmically large contributions;
however, we have verified that the leading vertex and self energy
corrections do not contain logarithms.

\subsection{Physical consequences}

We now discuss the physical content of the results.
We first note that the fermion-gauge field interaction has two effects on the
polarizability near $Q= 2p_F$: it changes the form of the nonanalyticity at
$\omega = 0$ and $Q= 2p_F$  (introducing the exponent $\sigma$) and it washes
out the nonanalyticity associated with the lower boundary of the particle-hole
continuum at $\omega = 2 v_F k_{\parallel}$.
Therefore the scaling form, eq. (\ref{chi_0}), gives the correct result for
the imaginary part of the susceptibility,
$\chi^{\prime\prime}$, which may be  measured in neutron scattering.
Using this and eq. (\ref{chi_RPA}) gives
\begin{equation}
\chi^{\prime\prime} (q, \omega ) =
\frac{\Pi^{\prime\prime} (q, \omega)}{(1 - g^2 \Pi(q,\omega)  )^2}
\end{equation}
The NMR $T_1$ relaxation rate is therefore
\begin{equation}
\frac{1}{T_1T} = A^2 g^4 \int dk_{\parallel} p_F d \theta
	\lim_{\omega \rightarrow 0}
		\frac{Im \Pi(k_{\parallel} , \omega)}{\omega}
	[\chi(0, k_{\parallel} ,\theta)]^2
\label{T1T_eq}
\end{equation}
Here $A$ is a constant proportional to the hyperfine coupling.
We can neglect the weak dependence of $\Pi$ on $\theta$ since
the singular dependence on $\theta$ comes only via $\chi$.
For $T > 0$ and $k_{\parallel}$ small ($k_{\parallel} \lesssim p_F
(T/J)^{2/3}$) we have
\begin{eqnarray}
\lim_{\omega \rightarrow 0} \frac{Im\Pi(k_\parallel=0,\omega)}{\omega} &=&
	\int G (\epsilon , p+Q/2) G(\epsilon ,p-Q/2)
 	[ \Gamma_{\epsilon ,p}^{(R)} (0,Q)]^2
	\frac{(dpd\epsilon )} {2T \cosh^2(\epsilon / 2T)} =
\nonumber \\
&=& c_T 	\sqrt{ \frac{p_0} {\omega_0 v_F^3}}
	\left( \frac{\omega_0}{T} \right)^{\frac{1}{3} + 2 \sigma}
\label{Im_Pi}
\end{eqnarray}
while for larger $k_\parallel$ it decreases as
\begin{equation}
\lim_{\omega \rightarrow 0} Im \Pi(k_\parallel,\omega) \sim Im \Pi (0, \omega )
\left( \frac{T^{2/3} \omega_0^{1/3}}{v_F |k_\parallel|}
	\right)^{(1+6\sigma )/2}
\end{equation}

The divergence of $Im \Pi (k_\parallel=0, \omega )/ \omega$ is the usual Kohn
anomaly modified by the fermion-gauge-field interaction.
This interaction has two effects.
First, the increased fermion damping proportional to $\epsilon^{2/3}$ weakens
the $T$ dependence from $T^{-1/2}$ to $T^{-1/3}$ but does not change the
$q$-dependence.
Second, the extra vertex correction then strengthens the $T$ dependence to
$T^{-(2\sigma + 1/3)}$ and the $q$-dependence to
$|q-2p_F|^{-(3 \sigma + 1/2)}$.
If $\sigma > 1/6$ as we assume throughout the strengthened $q$-dependence
leads to a divergence of $1/T_1 T\sim T^{(1-6 \sigma )/3}$ even far from
the critical point.

If the system is tuned to the critical point, then the $T$ dependence of
$\chi$ becomes important and we obtain
\begin{equation}
\frac{1}{T_1T} = A^2 \frac{1}{g^4}
	\sqrt{ \frac{v_F p_F}{\omega_0}}  \sqrt{ \frac{p_F}{p_0} }
	\left ( \frac{\omega_0}{T} \right )^{\frac{2}{3} - \sigma}
\label{T1T}
\end{equation}
If $\sigma < 1/3$, the critical contribution is more singular than the
background contribution.
However as $\sigma$ is increased to $1/3$, the difference between the
critical contribution and the background contribution disappears.

One may calculate the $T_2$ rate similarly.
The electronic contribution to the NMR $T_2^{-1}$ rate is related
to the real part, $\chi'$, of the susceptibility \cite{T2}:
\[
T_2^{-1} \sim
\frac{A^2}{a} \sqrt{\sum_k [ \chi^{\prime} ( \omega = 0 , k )]^2}
\]
where $a$ is the lattice constant.
If $\sigma < 1/9$, the critical contribution to $T_2^{-1}$ is
\[
T_2^{-1} \sim A^2 \frac{v_F}{a g^4} \sqrt{\frac{p_F}{p_0}}
	\left(\frac{\omega_0}{T} \right)^{1/6 - 3\sigma/2}
\]
Since we have assumed  $\sigma > 1/9$, $T_2^{-1}$ is not divergent as $T
\rightarrow 0$.

Proximity to the antiferromagnetic transition has also an effect on the
uniform susceptibility.
We have shown elsewhere \cite{chipaper} that the leading low-$T$ behavior of
$\chi({\bf q},0)$ in the limit
${\bf q} \rightarrow 0$
is given by
\begin{equation}
\chi_U = \lim_{{\bf q} \rightarrow 0} \chi({\bf q}, 0 ) =
\sum_{k,\nu} D (k,\nu) \chi(k,\nu)
\label{chi(0,0)_eq}
\end{equation}
where the coefficient $D$ is given by the diagram in Fig. 2b
with two of the spin fluctuation vertices replaced by vertices coupling the
fermions to the external magnetic field.
We denote this coupling to the external magnetic field by $g_e$.
In particular, terms arising from triangular vertices coupling two spin
fluctuations to one small-$q$ external field lead only to terms proportional
to integer powers of $T$, which are much smaller than the terms we keep.
The calculation of $D$ is very similar to the calculation of
$U$ (Eq. (\ref{U})) except that
the coupling $D$ has two small $q$ vertices and only two large
$Q$ vertices $\Gamma_{2p_F}$; the degree of singularity therefore differs
from that of $U$ and the dependence on $\theta_1 - \theta_2$ is absent.
We find
\begin{equation}
D (k_{\parallel} ,\nu) = \frac{p_0^{1/2}}{(v_F \omega_0)^{3/2}}
\frac{g^2 g_e^2}{\left ( \frac{\nu}{\omega_0} \right
)^{2/ 3 + 2 \sigma} + \left( \frac{k_{\parallel} v_F}{\omega_0}
	\right)^{3 \sigma +1}}
\label{D}
\end{equation}

Substituting this expression for $D$ and Eq. (\ref{chi_0}) for
$\chi$ into Eq. (2.13) we find at $g = g_c$ that
\begin{equation}
\chi_U = const + D_0 (T/ \omega_0 )^{\frac{2}{3} - \sigma}
\label{chi(0,0)}
\end{equation}
where $D_0$ is a constant of order $g_e^2 p_F/(g^2 v_F)$; $D_0$ is
positive if $\kappa/p_F > (T/\omega_0)^{1/3-\sigma}$ and $D_0$ is negative if
$\kappa/p_F < (T/\omega_0)^{1/3-\sigma}$.
The relation between the sign of $D_0$ and the magnitude of $\kappa$
comes because there are two sources of $T$-dependence in Eq.
(\ref{chi(0,0)_eq}): one is the $T$-dependence of the cut-off $\kappa$;  the
other is the discreteness of the frequency variable.

If $\kappa (T=0) >0$ then a crossover occurs.
For $T^{\frac{1}{3} - \sigma} > \kappa$, the result (\ref{chi(0,0)})
holds.
For lower $T$, the renormalization of the $2p_F$ interaction by the gauge
field shown in Fig. 6 becomes important and the $T$ dependence becomes weaker.
We discuss the $T$-dependence in the $g < g_c$ regime in the next section
because it is important for $\sigma > 1/3$.

\begin{figure}
\centerline{\epsfxsize=6cm \epsfbox{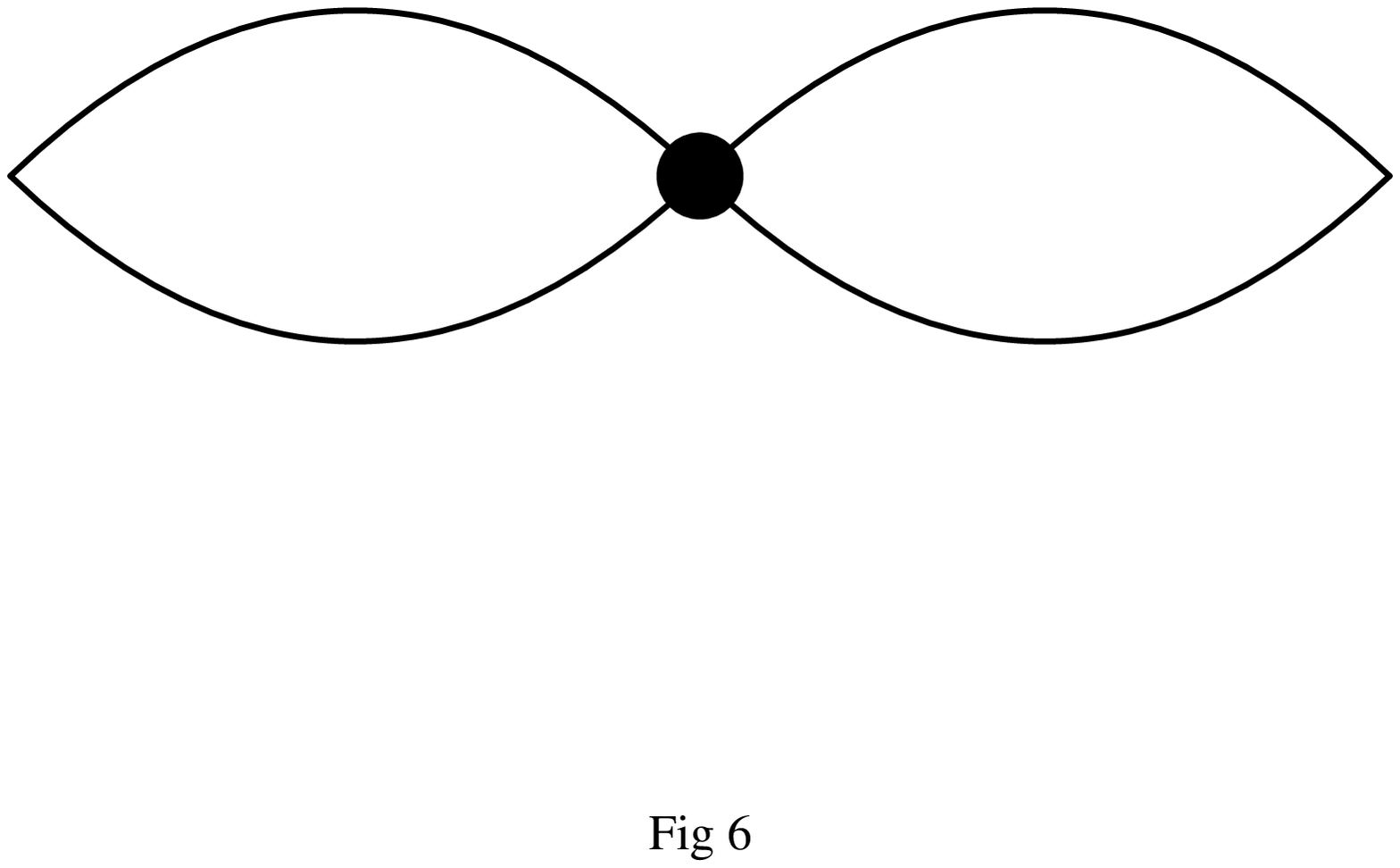}}
\caption{
Diagram yielding leading $T$-dependence of the uniform susceptibility for the
case of strong
gauge-field-fermion coupling.  The solid lines are fermion propagators and the
heavy dot is the short-range four fermion interaction dressed by the gauge
field.
}
\end{figure}

\section{Strong gauge field interaction $\sigma  > 1/3$}

If the exponent $\sigma$ arising from the fermion-gauge-field interaction is
greater than the critical value $1/3$, then the fermion spin susceptibility
diverges at $q=2p_F$ and $\omega =0$ for all $g < g_c$, leading to a phase
with power-law spin correlations at $T=0$.
At $g=g_c$ a transition will occur to a phase with long range order.
We have not yet formulated a theory of this transition.
In this section we outline the physical  consequences expected for the $g <
g_c$ phase.

For $\sigma > 1/3$ the Kohn singularity at $q=2p_F$ becomes nonintegrable, so
$\chi^{\prime}$ and $\chi^{\prime\prime} / \omega$ diverge at $T=0$ as shown
in Eq. (\ref{Pi2}).
The power law describing the divergence is $\theta$-independent.
The NMR relaxation rates are therefore given by summing the appropriate
combinations of susceptibilities over $k_{\parallel}$.
We find
\begin{equation}
\frac{1}{T_1T} \sim A^2 \frac{p_F p_0^{1/2} \omega_0^{1/2}}{v_F^{5/2}}
	\left( \frac{T}{\omega_0} \right)^{\frac{1}{3} - 2 \sigma}
\label{T1T_SL_1}
\end{equation}
If $\sigma < 1/2$, the rate $T_2^{-1}$ is non divergent and if $\sigma > 1/2$,
\begin{equation}
\frac{1}{T_2}  \sim A^2 \frac{\sqrt{p_F p_0} \omega_0}{v_F^2 a}
	\left( \frac{T}{\omega_0} \right)^{1-2 \sigma}
\label{T2_SL_1}
\end{equation}
Here $a$ is the lattice constant.
We now consider the uniform susceptibility $\chi(q \rightarrow 0 ,0)$.
It has contributions from diagrams involving the four fermion vertex $W$.
The leading diagram is shown in Fig. 6.
Because, as we have shown in a previous paper \cite{AIM}, $W$ is renormalized
by the gauge-field, $\chi_U$ acquires an anomalous temperature dependence.
The renormalization of $W$ is cut off by the largest of the temperature, the
energy of any fermion line, the scaled momentum $(vp_{\parallel})^{3/2}$ of
any fermion line, or the difference from $\pi$ of the angle
$\theta_1-\theta_2$ between the momentum of the incoming particles, leading to
\cite{AIM}
\begin{equation}
W \sim W_0 \max \left [
\frac{\omega}{\omega_0} , \left(\frac{k_{\parallel} v_F}{\omega_0}
	\right)^{3/2},
	(\theta_1 - \theta_2 - \pi)^3 \right ]^{\beta}
\label{W}
\end{equation}
where the exponent $\beta$ is $N$-dependent and is not simply related to
$\sigma$.
As $N \rightarrow \infty$, $\beta (N) = \frac{4}{3} - \frac{1}{N}$;
as $N \rightarrow 0$ $\beta(N) \sim c_1 \sqrt{N}$ with $c_1>0$.

To calculate the temperature dependence of $\chi(q\rightarrow 0,0)$ we insert
the vertex given in eq. (\ref{W}) into Fig. 6, and then determine the phase
volume in the $(\epsilon ,p)$, $(\epsilon^{\prime}, p^{\prime} )$ integrals
in which the $T$-dependence of $W$ is important.
We find
\begin{equation}
\chi_U = const +
D_0^{\prime \prime} (T/ \omega_0)^{1+\beta}
\end{equation}
where $D_0''$ is a constant of order $\left( \frac{p_F g_e}{v_F} \right)^2 W$,
whose sign is positive for repulsive $W$ and negative for attractive $W$.

\section{Conclusion}

We have determined the scaling behavior of the spin susceptibility near an
antiferromagnetic critical point at a ``$2p_F$'' wavevector of a
spin liquid.
By a ``$2p_F$'' wavevector we mean one which connects two points on the Fermi
surface with parallel tangents.
An example is shown in Fig. 1.
We distinguish between very weak, weak and strong fermion-gauge-field
coupling.
The cases are defined by the value of the exponent $\sigma$ appearing in Eq.
(\ref{Gamma_2pF}).
This exponent depends only on the fermion spin degeneracy, $N$; we do not know
how to calculate it analytically for the physically relevant case $N=2$.
For estimates of $\sigma$, see ref. \cite{AIM}.
In the very weak coupling case, $\sigma < 1/6$, a second order transition
at $|{\bf Q}|=2p_F$ is impossible because the maximum of $\chi$ is at
$|{\bf Q}|  \neq 2p_F$.
In the weak coupling case $1/6<\sigma<1/3$ the critical point occurs at $T=0$
when a short-ranged interaction $W$ equals a critical value $W_c$.
In the strong coupling case $\sigma >1/3$ there is a critical phase with power
law correlations at $T=0$ for $0 \leq W < W_c$ and an additional transition,
which we did not study, at $W=W_c$.

In weak and strong coupling cases the $2p_F$ singularities of the fermions
lead to physically important effects.
First, the scaling is anisotropic.
The dependence of the susceptibility on wavevectors $k_{\parallel}$ which
are parallel to the ordering wavevector $\bf Q$ involves a different exponent
than does the dependence on wavevectors $k_{\perp}$ perpendicular to $\bf Q$.
The $k_{\perp}$ exponent takes the conventional Ornstein-Zernike value 2 while
the
$k_{\parallel}$ exponent is always between $0$ and $1$ (cf. eq. (\ref{chi_0})).
The small value of the $k_{\parallel}$ exponent comes from a long range
(in position space) interaction due to the $2p_F$ singularity of the fermions.
It drastically weakens the singularities associated with the transition.
For example, in the weakly coupled spin liquid case
the NMR $T_2$  relaxation rate, which is given by
$T_2^{-2} \sim \sum_q (\chi_q^{\prime} )^2$, does not diverge as $T
\rightarrow 0$, in contrast to transitions in $2D$ insulating magnets, where
$T_2^{-1} \sim 1/T$.
The $T_1^{-1}$ rate is weakly divergent, see Eq. (\ref{T1T}).
The uniform susceptibility is nonvanishing as $T \rightarrow 0$, but the
leading temperature dependence is a power of $T$ between $0$ and $1$; see
Eq. (\ref{chi(0,0)}).

In the strongly coupled spin liquid the scaling is again anisotropic.
There is no singular dependence of $\chi ({\bf q},0 )$ on the
direction of $\bf q$.
The only singular dependence is on the difference between $\bf q$ and ``$2{\bf
p}_F ( \theta)$'', the vector spanning the Fermi surface and parallel to $q$.
For $\sigma > 1/3$ the NMR relaxation rate $(T_1T)^{-1}$ is always divergent
as $T \rightarrow 0$ (cf. Eq. (\ref{T1T_SL_1})); $T_2^{-1}$ diverges
if $\sigma > 1/2$ (cf. Eq. (\ref{T2_SL_1})).
Both quantities diverge more strongly as $\sigma$ is increased.
We also found that the leading temperature dependence of $\chi$ is $\chi \sim
const + T^{1+ \beta}$ where $\beta$ is an independent exponent which tends to
$0$ as $\sigma$ becomes large and to $4/3 - 2 \sigma$ as $\sigma \rightarrow
0$ (cf. Eq. (\ref{chi(0,0)})).
Our results for $T_1$, $T_2$ and the uniform susceptibility are summarized in
the Table.

One reason for studying spin fluctuations in two dimensional systems is that
the high-$T_c$ superconductors have been  shown to have strong
antiferromagnetic spin fluctuations.
The two best studied high-$T_c$ materials are $La_{2-x}Sr_xCuO_4$ and
$YBa_2Cu_3O_{7-\delta}$.
In $La_{2.x}Sr_xCuO_4$, neutron scattering has observed peaks centered at
incommensurate $x$-dependent wavevectors \cite{Aeppli}.
At $x=0.14$ the incommensurability was shown to correspond to a ``$2p_F$''
vector of the LDA bandstructure \cite{Littlewood}, suggesting that the results
of the present paper should be relevant.
NMR experiments have shown that the copper $T_1$ rate $^{Cu}T_1^{-1}$ has the
temperature dependence
\begin{equation}
^{Cu}(T_1T)^{-1} \sim 1/T
\end{equation}
for $100K < T < 500K$ \cite{Kitaoka}.
The $T_2$ rate has not been measured in this material,
but in other high $T_c$ materials with
($^{Cu}T_1T)^{-1} \sim 1/T$, $ T_2 \sim T^{-x}$ with
$\frac{1}{2} \lesssim x \lesssim 1$ \cite{Itoh}.
The uniform susceptibility is given by $\chi \sim const + AT$ at least for
$150K <T < 400K$ \cite{MillisMonien}.
None of these properties are consistent with
the weakly coupled spin liquid results of Section II.
The strongly coupled spin liquid results with $\sigma \approx 2/3$ are in
rough agreement with the data.

In the $YBa_2Cu_3O_{7-\delta}$ materials, $^{Cu}(T_1T)^{-1} \sim 1/T$
and $T_2^{-1} \sim T^{-x}$ for $T$ greater than a $\delta$-dependent
``spin-gap'' temperature and above this temperature
$\chi \sim const +A^{\prime} T$ with $A^{\prime}$ $\delta$-dependent.
The neutron scattering indicates broad and flat-topped peaks centered at the
commensurate wavevector $(\pi , \pi)$.
It is possible that the observed structure is due to several overlapping and
unresolved singularities at a $2p_F$ wavevectors.
However, it has also been argued
\cite{Levin} that in $YBa_2Cu_3O_{7-\delta}$ the magnetism is not driven by a
$2p_F$ instability, and is peaked at a commensurate wavevector because of a
strongly peaked interaction.
If the latter point of view is correct, the theory developed here is
irrelevant.
If the former point of view is correct, then we are forced to conclude that
the relevant fixed point is the strongly coupled spin liquid with $\sigma
\approx 2/3$.

One other aspect of the strongly coupled spin liquid requires further comment.
We noted already that in this case $\chi$ is singular at any
$Q$ spanning the Fermi surface.
In a translationally invariant system this would predict
peaks in the neutron scattering cross section  on a ring of radius $2p_F$
centered at the origin.
As noted by Littlewood et al \cite{Littlewood}, for fermions on a lattice one
obtains instead one or more curves traced out by the vectors $Q$ connecting
points with parallel tangents, and also one obtains additional families
of curves displaced by reciprocal lattice vectors, ${\bf G}$.
One gets further peaking when members of different families of curves
intersect.
The resulting structure is very sensitive to the details of the Fermi
surface and may resemble the data in $YBa_2Cu_3O_{7-\delta}$ as well as the
data on $La_{2-x}Sr_xCuO_4$.

We acknowledge the support of NSF grant DMR 90-22933.

\pagebreak

\begin{table}
\caption{Relaxation rates and susceptibility in different cases}
\begin{tabular}{cccc}
	& Weakly Coupled & Strongly Coupled \\
	&  Spin Liquid ($1/6 \!<\! \sigma \!<\! 1/3$)   &  Spin Liquid
						($\sigma \!>\! 1/3$) \\
\tableline 
$\frac{1}{T_1T}$ & $\frac{1}{T^{2/3-\sigma}}$ & $\frac{1}{T^{2\sigma-1/3}}$ \\
$\frac{1}{T_2}$ & non-divergent  &  $\frac{1}{T^{2\sigma-1}}$  \\
$\chi(q \rightarrow 0)$  &  $C + T^{2/3-\sigma}$  &  $C + T^{1+\beta}$
\end{tabular}
\end{table}


\begin{references}
\bibitem{remark}
The $2p_F$ wavevectors were referred to in previous work \cite{Millis} as
``maximal spanning vectors'' but this terminology is misleading because the
essential feature (parallel tangents) can occur even if $\bf Q$ is not a
maximal spanning vector.
\end{references}
\end{document}